\documentclass[11pt]{article}

\usepackage{a4,amssymb,verbatim,amsmath}
\newtheorem{ex}{Example}[section]
\newenvironment{example}{\begin{ex}\rm}{ \hfill $\Diamond$ \end{ex}
        \vskip4pt}
\newtheorem{ass}{Assumption}[section]

\numberwithin{equation}{section}

\begin{document}


\begin{center}
{\bf {\Large  One-dimensional gas dynamics equations
of  a polytropic  gas in Lagrangian coordinates:} \\
{\Large symmetry classification, conservation laws, difference
schemes}  }
\end{center}

\begin{center}
{\large Vladimir~A.~Dorodnitsyn$^{a}$, Roman~Kozlov$^{b}$, Sergey~V.~Meleshko$^{c}$}

\bigskip

$^{a}$
{Keldysh Institute of Applied Mathematics,
Russian Academy of Science,
     Miusskaya Pl. 4, Moscow, 125047, Russia}; \\
e-mail:  {Dorodnitsyn@Keldysh.com}

\medskip

$^{b}$
{Department of Business and Management Science,
Norwegian School of Economics,
Helleveien 30, 5045, Bergen, Norway}; \\
e-mail:  {Roman.Kozlov@nhh.no}

\medskip

$^{c}$
{School of Mathematics, Institute of Science,
       Suranaree University of Technology,
30000, Thailand}; \\
e-mail: {sergey@math.sut.ac.th}
\end{center}








\begin{abstract}
Lie point symmetries of the one-dimensional gas dynamics equations
of  a polytropic   gas in Lagrangian coordinates are considered.
Complete Lie group classification of these equations reduced to a
scalar second-order PDE is performed. 
The classification parameter is the entropy. Noether's
theorem is applied for constructing  conservation laws. The
conservation laws can be represented in the gas dynamics variables.
For the basic adiabatic case invariant and conservative
difference schemes are discussed.
\end{abstract}




\bigskip

Keyword:
Lagrangian gas dynamics,
Lie point symmetries,
Lie group classification,
conservation law,
Noether's theorem,
numerical scheme.

\bigskip











\section{Introduction}

There is an extensive literature on the Lagrangian gas dynamics
(see, for example, \cite{ bk:DespresMazeran2005, bk:Webb2018, bk:WebbZank[2009]}
and references therein). These studies can be roughly separated
on analysis of  the gas dynamics equations in the mass Lagrangian
coordinates \cite{bk:RozhdYanenko[1978], bk:DespresMazeran2005}
and
analysis of second-order PDEs to which these gas dynamics equations can be reduced
 \cite{bk:WebbZank[2009], bk:Webb2018,
bk:SiriwatKaewmaneeMeleshko2016, bk:VorakaKaewmaneeMeleshko2019}.
These second-order PDEs have variational structure,
i.e. they are  Euler-Lagrange equations for some Lagrangian function.
They are knows as the Lagrangian gas dynamics \cite{bk:DespresMazeran2005}.
In the present paper these Euler-Lagrange equations,
which represent   gas dynamics equations  in mass Lagrangian coordinates,
are studied.


Symmetries have always attracted the attention of scientists. One of
the tools for studying symmetries is the group analysis method
\cite{bk:Ovsiannikov1978,  bk:Olver[1986],  bk:MarsdenRatiu[1994],  bk:Ibragimov[1999], bk:Cantwell[2002]},
which is a basic method for constructing exact invariant solutions
of partial differential equations. The group properties of the gas
dynamics equations in Eulerian coordinates were studied in
\cite{bk:Ovsiannikov1978,bk:Ovsiannikov[1994]}.
Extensive group analysis of the one-dimensional gas dynamics equations in mass
Lagrangian coordinates was given in
\cite{bk:AkhatovGazizovIbragimov[1991],  Ames, bk:HandbookLie_v2}.
Here it should be also mentioned the results of \cite{bk:SjobergMahomed2004},
where nonlocal conservation laws of the one-dimensional gas
dynamics equations in mass Lagrangian coordinates were found. The
authors of \cite{bk:WebbZank[2009],bk:Webb2018} analyzed the
Euler-Lagrange equations corresponding to the one-dimensional gas
dynamics equations in mass Lagrangian coordinates: extensions of the
known conservation laws were derived. These conservation laws
correspond to special forms of the entropy. Group nature of these
conservation laws is given in the present paper. Complete group
analysis of the Euler-Lagrange equations of the one-dimensional gas
dynamics equations of isentropic flows of a polytropic
gas\footnote{The hyperbolic shallow water equations are equivalent
to the gas dynamics equations for a polytropic gas with
$\gamma=2$.} with $\gamma=2$ was given in
\cite{bk:SiriwatKaewmaneeMeleshko2016}.

As mentioned above, besides assisting with the construction of exact
solutions, the knowledge of an admitted Lie group allows one to derive
conservation laws. Conservation laws provide information on the basic
properties of solutions of differential equations. 
They are also needed in the analyses of stability and global behavior of solutions.
Noether's theorem \cite{bk:Noether[1918]} is the tool which relates
symmetries and conservation laws. However, an application of Noether's
theorem depends on the following condition: the differential equations
under consideration need to be presented  as Euler-Lagrange equations with
appropriate Lagrangian. Among approaches which try to overcome this
limitation one can mention here the approaches developed in
\cite{bk:Shmyglevski,Ibragimov2[2007],bk:BlumanCheviakovAnco,%
bk:SeligerWhitham[1968]}\footnote{Therein one can find more details and references.}  .


Lie group analysis of difference equations is a very active field of research
\cite{Levi2006, Winternitz2011, Dorodnitsyn2011}.
In our approach, which we are following in this paper,
we discretize differential equations while preserving their  Lie point symmetries.
 Thus differential equations and its Lie group symmetries  are a priory given
but not a difference model. One then looks for a difference scheme,
i.e. difference equations and a mesh, that have the same symmetry
group and the same Lie algebra. The basic steps in this direction
were done
\cite{Dorodnitsyn1989,
 Dorodnitsyn1993b,
BDK1997,
DKW2000,
DK2003,
DKW2004}, which were summarized in the book
\cite{Dorodnitsyn2011}.
The main
idea is that the invariant difference equations and meshes can be constructed with the
help of the entire set of difference invariants of the corresponding Lie group.


The present paper is focused on the one-dimensional gas dynamics
equations of  a polytropic gas in Lagrange coordinates. 
The objective of the paper is to make Lie group
classification of the variational second-order PDE to which the gas
dynamics equations can be reduced. The classification is carried out
with respect to the entropy function. Then, we use variational and
divergent symmetries to construct conservation laws by applying
Noether's theorem. 
In the second part of the paper  we consider discrete schemes
which preserve symmetries and conservation laws 
for the basic adiabatic case of the polytropic gas.


The paper is organized as follows:
Section  \ref{Gas_dynamics_equations} introduces
equations of gas dynamics, their variational formulation and Lie point symmetries.
In section \ref{Conservation_laws_kernel} we consider
the general case and the corresponding  conservation laws.
Then,  we present the cases with additional symmetries and conservation laws
in section \ref{Additional_conservation_laws}.
Discrete schemes for  the general case are discussed in section \ref{Finite_difference}.
Finally, section \ref{Concluding_remarks} gives concluding remarks.

\section{Gas dynamics equations}

\label{Gas_dynamics_equations}

In this section we present the one-dimensional  gas dynamics equations
for a polytropic gas.
We start with Euler coordinates and show how to change to Lagrangian ones.
It is shown how one can reduce the gas dynamics equations to a scalar second-order PDE,
which has variational structure.

\subsection{The gas dynamics equations in Eulerian coordinates}

The gas dynamics equations of a polytropic gas are
\begin{subequations}   \label{Euler_GD}
\begin{gather}
\rho_{t}+u\rho_{x}+\rho u_{x}=0,
\label{Euler_GD_rho}  \\
\rho(u_{t}+uu_{x})+p_{x}=0,
\label{Euler_GD_u}  \\
S_{t}+u S_{x}=0,
\label{Euler_GD_S}
\end{gather}
\end{subequations}
where $\rho$ is the density, $u$ is the velocity  and
\begin{equation}        \label{p_S_rho}
p=S \rho^{\gamma} ,
\qquad
\gamma \neq 0, 1 ,
\end{equation}
is the pressure. 
Here  $S$ is the function depending on the entropy $\tilde{S}$ of the gas.
For a polytropic gas this relation takes the form   \cite{bk:Ovsiannikov[2003]}
\begin{equation}
S=R  \   e^{(\tilde{S}-\tilde{S}_{0})/c_{v}}  ,
\end{equation}
where $R$ is the gas constant, $c_{v}$ is the dimensionless specific
heat capacity at constant volume and $\tilde{S}_{0}$ is constant.

Note that due to  (\ref{p_S_rho}) the last equation   (\ref{Euler_GD_S}) is equivalent to 
\begin{equation}
\label{Euler_GD_p}
p _{t}+ u p_{x}+  \gamma p  u_{x}=0   .
\end{equation}

\subsection{Eulerian and Lagrangian coordinates}

For one-dimensional gas dynamics one can introduce  mass Lagrangian   coordinate $s$
in Eulerian coordinates $ (t,x) $  as
\begin{equation}\label{Change}
 ds=\rho dx - \rho u dt ,
\end{equation}
or
\begin{equation}\label{Change2}
 s_x =  \rho ,
\qquad
 s_t  = -\rho u .
\end{equation}
We also need the time differentiation  in Lagrangian coordinates $(t,s)$,
which is called the material derivative,
\begin{equation}\label{Lagr-time}
D_t  ^L  = D_t  ^E + u D_x ,
\end{equation}
where     $D_t ^E $ and $ D_x $ are the total differentiations with respect  to $t$  and $x$ in Euler coordinates.
Thus we arrive at the gas dynamics equations
in Lagrangian coordinates:
\begin{subequations}   \label{Lagrange_GD}
\begin{gather}
   \left(   { 1 \over \rho }   \right)  _{t}  =   u_{s} ,
\label{Lagrange_GD_rho}   \\
  u_{t}+p_{s}=0,
 \label{Lagrange_GD_u}   \\
S_{t}=0 .
\label{Lagrange_GD_S}
\end{gather}
\end{subequations}
Note that the  equation   (\ref{Euler_GD_p}) takes the form
\begin{equation}
\label{Lagrange_GD_p}
p_{t} + \gamma  \rho p u_{s} = 0 .
\end{equation}

We remark that we use  a non-point change of the independent variables.

\subsubsection{Variational formulation }

Let us introduce  variable $\varphi$ as the potential for equation   (\ref{Lagrange_GD_rho}):
\begin{equation}      \label{potential} 
\varphi_{t}  = u ,
\qquad
\varphi_{s}  = { 1 \over \rho}
\end{equation}
and solve equation  (\ref{Lagrange_GD_S}) as
\begin{equation}
S = S(s)  .
\end{equation}
Then, with the help of (\ref{p_S_rho}) we can rewrite the remaining
equation (\ref{Lagrange_GD_u}) as the following second-order PDE \cite{bk:Chernyi_gas}:
\begin{equation}     \label{eq:one-D-gas_Lagrange}
\varphi_{s}^{\gamma}\varphi_{tt}  -  \gamma S\varphi_{s}^{-1}\varphi_{ss}  +  S_{s}=0     .
\end{equation}
This equation  will be called the gas dynamics  equation in  Lagrangian coordinates.
It  is the variational equation for the Lagrangian
\begin{equation}   \label{Lagrangian_function}
{L}=\frac{\varphi_{t}^{2}}{2}   -   \frac{  S(s)  }{\gamma-1}     \varphi_{s}^{1 -\gamma}  .
\end{equation}
Let us recall that the  Euler-Lagrange equation for Lagrangian  (\ref{Lagrangian_function})   is computed as
\begin{equation}
\frac{\delta{L}}{\delta\varphi}
= { \partial  L \over \partial \varphi }
-  D_t  ^L  \left( { \partial  L \over \partial \varphi_t }     \right)
-  D_s   \left( { \partial  L \over \partial \varphi_s }   \right)
 =    0  .
\end{equation}


A symmetry
\begin{equation}
X =\xi^{t} (t,s,\varphi)  \frac{\partial}{\partial t}
+\xi^{s} (t,s,\varphi) \frac{\partial}{\partial s}
+\eta^{\varphi} (t,s,\varphi) \frac{\partial}{\partial\varphi}
\end{equation}
of equation (\ref{eq:one-D-gas_Lagrange})  
provides its  conservation law 
if it satisfies the equation
\begin{equation}
   X  {L}+{L}(D_{t} ^L \xi^{t}+D_{s}\xi^{s})
=D_{t}  ^L B _{1}+D_{s}  B _{2}  ,
\label{eq:Noether}
\end{equation}
for some functions
$ B _{i}(t,s,\varphi) $, $ i=1,2$.
Here  we assume that  the operator   $X$ is prolonged on the derivatives
by means of the standard prolongation formulae  \cite{bk:Ovsiannikov1978, bk:Olver[1986]}.

The  densities $(T^{t}  ,T^{s})$ of the conservation laws 
are given by the formulae
\begin{equation}       \label{conservation_laws_densities}
T^{t}
=\xi  ^{t} {L}
+  (  \eta^{\varphi} - \xi^{t}\varphi_{t} - \xi ^{s}\varphi_{s}  )  \frac{\partial  L }{\partial   \varphi_{t} }  ,
\qquad
T^{s}
=\xi  ^{s}{L}
+  (  \eta^{\varphi} - \xi^{t}\varphi_{t} - \xi ^{s}\varphi_{s}  )  \frac{\partial  L }{\partial  \varphi_{s}}   .
\end{equation}

\subsubsection{Relation between conservation laws in Lagrangian and Eulerian coordinates}

The operators of the total differentiation  $D_{t} ^L $  and   $D_{s}$
in Lagrangian coordinates $(t,s)$
and the operators of the total differentiation  $D_{t} ^E $  and   $D_{x}$ 
in Eulerian coordinates  $(t,x)$
are related as follows
\begin{equation}
D_{t} ^L = D_{t}  ^E+ \varphi_{t}D_{x} = D_{t}  ^E+ u D_{x}   , 
\qquad 
D_{s}=\varphi_{s} D_{x} =\frac{1}{\rho}D_{x} . 
\end{equation}

Let $(T^{t}  ,T^{s})$ be a conserved vector in Lagrangian coordinates:
\begin{equation}
D_{t} ^L T^{t}  +  D_{s}T^{s}=0.
\end{equation}
Using    (\ref{potential}), we obtain  
\begin{equation}
u_{x}= \varphi_{s}^{-1}    \varphi_{ts}    
\end{equation}
and
\begin{equation}
D_{t} ^L T^{t}+D_{s}T^{s}
=  \varphi_{s} \left(    D_{t}  ^E (\rho T^{t})    +    D_{x}(     \rho uT^{t}+T^{s} )    \right).
\end{equation}
Thus, the corresponding  conserved vector $(^{e}T^{t}  ,{}^{e}T^{x})$ in Eulerian
coordinates has the coordinates
\begin{equation}     \label{transformation_rule}
^{e}T^{t}=\rho T^{t}  ,
\ \
{}^{e}T^{x}=\rho uT^{t}+T^{s}  .
\end{equation}

\subsection{Equivalence transformations}

The group classification of differential equations depends on representations of
arbitrary elements \cite{bk:Ovsiannikov1978}.
As the function $S(s)$ is an arbitrary element of equation (\ref{eq:one-D-gas_Lagrange}),
the group classification has to be made with respect to it.

As the first step in the symmetry group classification
we find equivalence Lie group transformations.
The generator of the equivalence transformations
is considered in the form \cite{bk:Meleshko[2005]}
\begin{equation}
X^{e} = \xi^{t} \frac{\partial}{\partial t} +\xi^{s}
\frac{\partial}{\partial s} +\eta^{\varphi}
\frac{\partial}{\partial\varphi} +\eta^{S} \frac{\partial}{\partial
S}  ,
\end{equation}
where all the coefficients of the generator $X^{e}$ depend on $(t,s,\varphi,S)$.

Calculations show that the Lie algebra corresponding to the equivalence
Lie group of equation (\ref{eq:one-D-gas_Lagrange}) consists of the
generators
\begin{multline}      \label{equivalence_general}
X_{1}^{e}=\frac{\partial}{\partial\varphi} ,
\qquad
X_{2}^{e}=  \frac{\partial}{\partial t} ,
\qquad
X_{3}^{e}=   \frac{\partial}{\partial s}  ,
\qquad
X_{4}^{e}=t  \frac{\partial}{\partial\varphi} ,
\\
X_{5}^{e}
= \varphi    \frac{\partial}{\partial\varphi}
+ (\gamma+1) S \frac{\partial}{\partial  S } ,
\qquad
X_{6}^{e}
= t  \frac{\partial}{\partial t }
-2S  \frac{\partial}{\partial  S }  ,
\\
X_{7}^{e}
=s  \frac{\partial}{\partial  s }
+(1-\gamma)S  \frac{\partial}{\partial  S }  .
\end{multline}
For $\gamma=3$ there are more equivalence transformations corresponding
to the generator
\begin{equation}      \label{equivalence_special}
X_{*}^{e}= t^2  \frac{\partial}{\partial t }
+  t \varphi    \frac{\partial}{\partial\varphi}  .
\end{equation}

\subsection{Group classification of equation (\ref{eq:one-D-gas_Lagrange})}

Group classification of equation (\ref{eq:one-D-gas_Lagrange})
was given in \cite{bk:AndrKapPukhRod[1998]}.
For the sake of completeness we give it here.
The admitted generator has the form
\begin{equation}      \label{symmetry_form}
X  =  \sum_{i=1}^{8}k_{i}  Y _{i}  ,
\end{equation}
where
\begin{multline}
Y_{1}=\varphi\frac{\partial}{\partial\varphi}  ,
\qquad
Y_{2}=  t ^2 \frac{\partial}{\partial t}
+ t \varphi  \frac{\partial}{\partial\varphi} ,
\qquad
Y_{3}=t\frac{\partial}{\partial t}  ,
\qquad
Y_{4}=\frac{\partial}{\partial t}  ,
\\
Y_{5}=t\frac{\partial}{\partial\varphi}  ,
\qquad
Y_{6}=\frac{\partial}{\partial\varphi}  ,
\qquad
Y_{7}=s\frac{\partial}{\partial s}  ,
\qquad
Y_{8}=\frac{\partial}{\partial s}  .
\end{multline}
The coefficients  $k_i$ satisfy the conditions
\begin{subequations}   \label{conditions}
\begin{gather}
(k_{7}s+k_{8})  S_{s}
=   (   (1-\gamma)   k_{7}     +  (\gamma+1)    k_{1}  -  2k_{3}  )   S  ,
\label{condition1}
 \\
 (\gamma-3)   k_{2}   =  0  .
\label{condition2}
\end{gather}
\end{subequations}


Thus, in the generic case ($\gamma \neq 3 $)
the kernel of the algebra of the admitted Lie symmetries 
can be presented by the following basis
\begin{multline}        \label{kernel1}
X_1 = Y_{6}=\frac{\partial}{\partial   \varphi}  ,
\qquad
X_2 = Y_{4}=\frac{\partial}{\partial t}  ,
\qquad
X_3 = Y_{5}=t\frac{\partial}{\partial \varphi}  ,
\qquad
\\
X_4
= 2 Y_{1}+   ( \gamma+1 )  Y_{3}
=  (\gamma+1)  t\frac{\partial}{\partial t}
+  2 \varphi\frac{\partial}{\partial \varphi}   .
\end{multline}
For $\gamma = 3 $ the kernel of the admitted Lie symmetries algebra  has one additional generator
\begin{equation}       \label{kernel2}
X_*  = Y_{2}
=  t ^2 \frac{\partial}{\partial t}
+ t \varphi  \frac{\partial}{\partial \varphi}     .
\end{equation}

According to the classifying equation (\ref{condition1}),
there exists constants  $\alpha$, $\beta$ and $q$  such that
the entropy function satisfies the equation
\begin{equation}    \label{eq:Nov1.1}
(\alpha s+\beta)   S_{s} = q S .
\end{equation}
Let us consider possible cases:

\begin{enumerate}

\item

$ q = 0 $ ($\alpha \neq 0 $ or $\beta \neq 0 $)

I this case   equation   (\ref{eq:Nov1.1}) gives
\begin{equation}      \label{first_case}
S(s) =   A_0 ,
\qquad
A_0   = \mbox{const}   .
\end{equation}

\item

$ q  \neq 0 $

\begin{enumerate}

\item

If $\alpha\neq0$, then by virtue of the equivalence transformations
equation (\ref{eq:Nov1.1}) can be reduced to the equation
\begin{equation*}
sS_{s}=qS ,
\qquad
 q \neq 0 .
\end{equation*}
The general solution of the latter equation is
\begin{equation}       \label{second_case}
S (s) =A_{0} s^{q}  ,
\qquad
A_0   = \mbox{const}   .
\end{equation}

\item

If $\alpha=0$ and $\beta\neq0$,
  equation (\ref{eq:Nov1.1}) can be reduced to 
\begin{equation*}
S_{s}=q S,
\qquad
 q \neq 0 ,
\end{equation*}
which has the general solution
\begin{equation}      \label{third_case}
S = A_{0} e^{q s}  ,
\qquad
A_0   = \mbox{const}   .
\end{equation}

\end{enumerate}

\end{enumerate}

Thus, one needs to study three cases:
  (\ref{first_case}),    (\ref{second_case})   and   (\ref{third_case}).
Notice that the first case   (\ref{first_case})  corresponds to an  isentropic flow of a polytropic  gas.
It is  well-studied,
whereas the other cases  were not analyzed in the theory of gas dynamics equations.
We remark  that by virtue of the equivalence transformations one can assume that $A_{0}=1$.

\section{Conservation laws for arbitrary entropy  $S(s)$}

\label{Conservation_laws_kernel}

In  Lagrangian coordinates the conservation law of mass
in given by the first equation of the gas dynamics system (\ref{Lagrange_GD}),
namely
\begin{equation}   \label{conservation_mass}
\left( 1 \over \rho \right)_{t} - u_{s}=0  .
\end{equation}
For variational and divergent symmetries 
we can  using Noether's theorem to  derive conservation laws. 
In this section we exploit the generators from the kernel of admitted Lie symmetry algebra.

\subsection{Case $\gamma\protect\neq3$ }

\label{point_kernel_1}

Let us examine symmetries  (\ref{kernel1}) of the generic case   $ \gamma \neq 3$.
For Lagrangian     (\ref{Lagrangian_function})
we get  two variational symmetries
\begin{equation}     \label{Noether_symmetry_0_1}
Z_1 = X_{1}  =   { \partial \over \partial \varphi }  ,
\qquad
Z_2 = X_{2}  =   { \partial \over \partial  t }
\end{equation}
and  one divergent symmetry
\begin{equation}     \label{Noether_symmetry_0_2}
Z_3 = X_{3}   =  t  { \partial \over \partial \varphi }
 \quad
 \mbox{with}   \quad  ( B _1  , B_2 ) =  ( \varphi  , 0 )     .
\end{equation}
The corresponding conservation laws are given  by (\ref{conservation_laws_densities}):
\begin{equation}
T_{1}^{t}=\varphi_{t} ,
\qquad
T_{1}^{s}=S\varphi_{s}^{-\gamma}  ;
\end{equation}
\begin{equation}
T_{2}^{t}= { \varphi_{t}^{2}  \over 2 }
+ { S   \over  \gamma-1  }    \varphi_{s}^{1-\gamma} ,
\qquad
T_{2}^{s}= S\varphi_{t}\varphi_{s}^{-\gamma}   ;
\end{equation}
\begin{equation}
T_{3}^{t}
=  \varphi  -  \varphi_{t}t  ,
\qquad
T_{3}^{s}= -  t S \varphi_{s}^{-\gamma}     .
\end{equation}
They represent the conservation of momentum, the conservation of energy
and the motion of the center of mass, respectively.

The  remaining kernel symmetry  operator
$  X_4  $
is neither variational nor divergent.
Hence, it does not provide any conservation law.

These conservations laws can be rewritten  in the gas dynamics variables  as
\begin{equation}     \label{conservation_momentum}
T_{1}^{t}=u ,
\qquad
T_{1}^{s}=  S \rho^{\gamma}  ;
\end{equation}
\begin{equation}   \label{conservation_energy}
T_{2}^{t}
=   { u^{2}   \over 2 }
+ { S   \over   \gamma-1  }   \rho^{\gamma-1}  ,
\qquad
T_{2}^{s}
=   S \rho^{\gamma} u   ;
\end{equation}
\begin{equation}       \label{conservation_center_of_mass}
T_{3}^{t}=  \varphi   -  t u   ,
\qquad
T_{3}^{s}=-    t   S \rho^{\gamma}   .
\end{equation}
Notice that the conserved vector $(T_{3}^{t}  ,T_{3}^{s})$ contains
the function $\varphi$.

It is possible to  present these conservation laws  in Eulerian coordinates
applying  the  transformation rule  (\ref{transformation_rule}).
We get  the conservation law for momentum and  energy and the motion of the center of mass as
\begin{equation*}
{}^{e}T_{1}^{t}=\rho u ,
\qquad
{}^{e}T_{1}^{x}=\rho u^{2}+S\rho^{\gamma}  ;
\end{equation*}
\begin{equation*}
{}^{e}T_{2}^{t}
=    { \rho u^{2}  \over 2 }
+ {  S \rho^{\gamma}   \over  \gamma-1    }  ,
\qquad
{}^{e}T_{2}^{x}
= \left(  { \rho u^{2}  \over 2 }
+ {  \gamma S \rho^{\gamma}    \over  \gamma-1    }     \right) u  ;
\end{equation*}
\begin{equation*}
{}^{e}T_{3}^{t}=\rho( x - t u ) ,
\qquad
{}^{e}T_{3}^{x}=\rho u ( x - t u ) -  t S \rho^{\gamma}    .
\end{equation*}



\subsection{Case $\gamma=3$}

\label{point_kernel_2}

For  $\gamma=3$  we present only the additional conservation laws.
The generic kernel symmetry
\begin{equation}     \label{Noether_symmetry_0_3}
Z_*  =    X_4
=    2   t \frac{\partial}{\partial t}
+   \varphi  \frac{\partial}{\partial\varphi}
\end{equation}
becomes   variational.
We also have symmetry (\ref{kernel2}),
which  is   divergent
\begin{equation}     \label{Noether_symmetry_0_4}
Z_{**} = X_*
=  t  ^2 \frac{\partial}{\partial t}
+ t  \varphi  \frac{\partial}{\partial\varphi}
\qquad
\mbox{with}
\qquad
( B _1  , B_2 ) =  \left( { \varphi ^2  \over 2 }  , 0  \right)    .
\end{equation}

The application of Noether's theorem gives
the conservation laws of the generic case, which were given in the previous point,
and the following two additional conservation laws:
\begin{equation}
T_{*}^{t}
=
2  t
\left(  { \varphi_{t}  ^2    \over 2 }
+  { S \varphi_{s}^{-2}   \over \gamma - 1}
\right)
 -   \varphi   \varphi_{t}   ,
\qquad
T_{*}^{s}
=(  2 t \varphi_{t}  - \varphi  ) S \varphi_{s}^{-3}  ;
\end{equation}
\begin{equation}
T_{**}^{t}
=
  t^2    \left(
{   \varphi_{t} ^{2}   \over 2 }
+   {  S \varphi_{s}^{-2}     \over \gamma -1 }
\right)
-    t  \varphi   \varphi_{t}
+   { \varphi ^2  \over 2 }    ,
\qquad
T_{**}^{s}
=  (  t^2 \varphi_{t}    -  t  \varphi   ) S \varphi_{s}^{-3}   .
\end{equation}
In the gas dynamics variables these conservations laws take the form
\begin{equation}    \label{conservation_additional_1}
T_{*}^{t}
=
2  t
\left(  { u ^2    \over 2 }
+  { S \rho ^{2}   \over \gamma - 1}
\right)
 -   \varphi  u    ,
\qquad
T_{*}^{s}
=(  2 t u   - \varphi  ) S \rho ^{3}   ;
\end{equation}
\begin{equation}     \label{conservation_additional_2}
T_{**}^{t}
=
  t^2    \left(
{   u  ^{2}   \over 2 }
+   {  S \rho ^{2}     \over \gamma -1 }
\right)
-    t  \varphi   u
+   { \varphi ^2  \over 2 }    ,
\qquad
T_{**}^{s}
=  (  t^2  u    -  t  \varphi   ) S \rho ^{3}     .
\end{equation}
It should be noted that these  conserved vectors were presented in \cite{bk:SjobergMahomed2004},
where $\varphi$ was considered as a nonlocal variable.

It is also possible to transform these conservation laws into Eulerian coordinates:
\begin{equation*}
{}^{e}T_{*}^{t}
= 2  t \left(
  { \rho   u ^{2}  \over 2 }
+  { S \rho^{3}    \over \gamma - 1 }
\right)
 -  x   \rho      u   ,
\qquad
{}^{e}T_{*}^{x}
=
2t   \left(
{ \rho  u ^{2}  \over 2 }
+ {  \gamma  S \rho^{3}   \over \gamma - 1  }
\right) u
-  x  (  \rho u ^2   +     \rho^{3} S   )    ;
\end{equation*}
\begin{multline*}
{}^{e}T_{**}^{t}
=  t^2 \left(
  { \rho   u ^{2}  \over 2 }
+  { S \rho^{3}     \over \gamma - 1 }
\right)
 -   t x \rho   u
+  { x ^{2}  \over 2 } \rho   ,
\\
{}^{e}T_{**}^{x}
=
t^2 \left(
{ \rho  u ^{2}  \over 2 }
+ {  \gamma  S \rho^{3}   \over \gamma - 1  }
\right) u
- t x  (  \rho u ^2   +     \rho^{3} S   )
+  { x  ^{2}   \over 2 }  \rho u   .
\end{multline*}

\section{Conservation laws for special cases of  entropy  $S(s)$}

\label{Additional_conservation_laws}

There can be additional symmetries admitted by equation (\ref{eq:one-D-gas_Lagrange})
for some cases of the entropy function $S (s)$.
In this section we check whether these additional symmetries lead to additional conservation laws.

\subsection{Isentropic flow $S (s) =A_{0}$}


In Eulerian coordinates the isentropic case is selected
out by the differential constraint 
\begin{equation}
S_{x}=0 . 
\end{equation}

The admitted generator has the form (\ref{symmetry_form})
with the coefficients satisfying the equations
\begin{subequations}
\begin{gather}
    (1-\gamma)   k_{7}     +  (\gamma+1)    k_{1}  -  2k_{3}     = 0  ,     \\
  (\gamma-3 )   k_{2}    =  0  .
\end{gather}
\end{subequations}
In  both cases $\gamma\neq3$ and $\gamma=3$  we get
two  symmetries
\begin{equation}       \label{additional_symmetry_1_0}
X_{5}
=(\gamma-1)Y_{3}-2Y_{7}
=(\gamma-1)t\frac{\partial}{\partial t}-2s\frac{\partial}{\partial s}  ,
\qquad
X_{6}=    Y_8  =   \frac{\partial}{\partial s}
\end{equation}
in  addition to the symmetries  given in (\ref{kernel1}) and (\ref{kernel2}).
For application of Noether's theorem \cite{bk:Noether[1918]}
the study has to be split into the cases  $\gamma\neq3$ and $\gamma=3$.

\subsubsection{Case $\gamma\protect\neq3$ }


The admitted symmetries are given by  (\ref{kernel1})     and      (\ref{additional_symmetry_1_0}).
Checking  when  the symmetry generator
\begin{equation}
X =  \sum _{i  = 1}   ^6    \beta_i   X_i
\end{equation}
is variational or divergent by substitution into (\ref{eq:Noether}),
we get the condition
\begin{equation}
\beta _4 ( \gamma -3 ) + \beta _5 ( \gamma + 1 ) = 0 .
\end{equation}
Therefore, we get two variational symmetries
\begin{multline}
Z_{4} =  { \gamma+1 \over 2 } X_{4}  -   { \gamma - 3 \over 2 } X_{5}
= (3\gamma-1)t\frac{\partial}{\partial t}
+(\gamma-3)s\frac{\partial}{\partial s}
+(\gamma+1)\varphi\frac{\partial}{\partial\varphi}  ,
\\
Z_{5}=   X_6 = \frac{\partial}{\partial s}  ,
\end{multline}
in addition to symmetries  (\ref{Noether_symmetry_0_1})
and  (\ref{Noether_symmetry_0_2}).

Using Noether's  theorem,  we find conservation laws given in point \ref{point_kernel_1}
and two conservation laws, corresponding to the additional variational   symmetries:
\begin{multline}
T_{4}^{t}
=
   ( 3 \gamma - 1 )  t \left(
 { \varphi_{t}^{2}   \over 2 }
+ { A_{0}  \over \gamma -1}  \varphi_{s}^{1-\gamma}
\right)
+    (  \gamma - 3 )    s \varphi_{t}\varphi_{s}
-       (\gamma  + 1)      \varphi\varphi_{t}   ,
\\
T_{4}^{s}
=
  ( 3 \gamma - 1 ) t  A_{0}  \varphi_{t}     \varphi_{s}^{-\gamma}
+   (  \gamma - 3 )    s \left(
-    {  \varphi_{t}^{2} \over 2 }
+ {  \gamma   A_{0}  \over \gamma -1 }    \varphi_{s}^{1-\gamma}
\right)
-  (\gamma  + 1 )  A_{0} \varphi   \varphi_{s}^{-\gamma}    ;
\end{multline}
\begin{equation}
T_{5}^{t}
=    \varphi_{t}\varphi_{s}  ,
\qquad
T_{5}^{s}=  -   { \varphi_{t}^{2}  \over 2 }
 +   {   \gamma A_{0}    \over   \gamma-1  }       \varphi_{s}^{1-\gamma}  .
\end{equation}
These additional conservation laws take the form
\begin{multline}
T_{4}^{t}
=
   ( 3 \gamma - 1 )  t \left(
 { u ^{2}   \over 2 }
+ { A_{0}  \over \gamma -1}  \rho ^{\gamma-1}
\right)
+    (  \gamma - 3 )    s  { u \over \rho  }
-       (\gamma  + 1)      \varphi  u    ,
\\
T_{4}^{s}
=
  ( 3 \gamma - 1 ) t  A_{0}  u      \rho ^{\gamma}
+  (  \gamma - 3 )   s \left(
-     {  u ^{2} \over 2 }
+ {  \gamma    A_{0} \over \gamma -1 }      \rho ^{\gamma-1}
\right)
-  (\gamma  + 1 )  A_{0} \varphi   \rho ^{\gamma}  ;
\end{multline}
\begin{equation}
T_{5}^{t}
= {  u  \over  \rho  }   ,
\qquad
T_{5}^{s}
= -  { u^{2}  \over 2 }
+   {   \gamma A_{0}  \over   \gamma -1} \rho^{\gamma-1}
\end{equation}
in the gas dynamics variables.

In Eulerian coordinates the latter conservation law becomes
\begin{equation*}
{}^{e}T_{5}^{t}
=  u   ,
\qquad
{}^{e}T_{5}^{x}
= { u^{2} \over 2 }
+  {  \gamma    A_{0}   \over \gamma -1}   \rho^{\gamma-1}  .
\end{equation*}
The densities of the conserved vector
$(T_{4}^{t}  ,T_{4}^{s})$ contain the Lagrangian mass coordinate $s$.
For this reason they cannot be rewritten as a local conservation law in Eulerian coordinates.

\subsubsection{Case $\gamma=3$}


We have symmetries   (\ref{kernel1}),   (\ref{kernel2}) and (\ref{additional_symmetry_1_0}).
Substituting the generator
\begin{equation}
X = \sum  _{i=1} ^6   \beta _i  X_i    +    \beta _*   X_*
\end{equation}
into (\ref{eq:Noether}), we get the condition
\begin{equation}
 \beta _5 = 0   .
\end{equation}
Therefore we get only one conservation law
in addition to conservation laws given in points \ref{point_kernel_1} and    \ref{point_kernel_2}.
It corresponds  to the symmetry
\begin{equation}
Z _{4} = X_6  =\frac{\partial}{\partial s}    .
\end{equation}

The conserved vector has components 
\begin{equation}
T_{4}^{t}
=  \varphi_{t}\varphi_{s}  ,
\qquad
T_{4}^{s}
= -  { \varphi_{t}^{2}  \over 2 }
 + { \gamma  A_{0}  \over \gamma -1}   \varphi_{s}^{-2}   .
\end{equation}
In the gas dynamics variables it takes the form
\begin{equation}
T_{4}^{t}
=  { u \over \rho } ,
\qquad
T_{4}^{s}
= -{  u^{2} \over 2 }
+  {  \gamma A_{0}  \over \gamma -1} \rho^{2}  .
\end{equation}
In Eulerian coordinates we can present it as
\begin{equation*}
{}^{e}T_{4}^{t}= u,
\qquad
{}^{e}T_{4}^{x}
=  { u^{2}  \over 2 } +  { \gamma A_{0}  \over \gamma -1 }  \rho^{2}  .
\end{equation*}

\subsection{Entropy $S=A_{0}s^{q}$}


In Eulerian coordinates this case is described by the differential constraint
\begin{equation}
q\rho SS_{xx}-q\rho S_{x}^{2}-q\rho_{x}SS_{x}+\rho S_{x}^{2}=0.
\label{eq:dif_con2}
\end{equation}
One can check that the overdetermined system of equations 
consisting  of the gas dynamics equations and this  constraint is involutive.

Calculations show that the admitted generators have the form  (\ref{symmetry_form})
and the coefficients satisfy 
\begin{subequations}
\begin{gather}
     (1-\gamma  -q  )     k_{7}     +  (\gamma+1)    k_{1}  -  2k_{3}    = 0    ,     \\
k_{8}   = 0 ,   \\
 (\gamma-3)   k_{2}   =  0  .
\end{gather}
\end{subequations}
In both cases    $ \gamma \neq 3 $  and  $ \gamma = 3 $
we get only one additional generator
\begin{equation}       \label{additional_2_0}
X_{5} = (1-\gamma-q) Y_{3}  +  2 Y_{7}
=(1-\gamma-q)t\frac{\partial}{\partial t}+2s\frac{\partial}{\partial s}  .
\end{equation}


As in the previous  case of the entropy function  $S(s)$,
one needs to study {sub}cases $\gamma=3$ and $\gamma\neq3$ separately.


\subsubsection{Case $\gamma\protect\neq3$}

Substituting the generator
\begin{equation}
X = \sum  _{i=1} ^5   \beta _i  X_i
\end{equation}
into (\ref{eq:Noether}),
we find the condition for Noether (variational and divergent) symmetries
\begin{equation}
 ( 3 - \gamma )    \beta _4
+  ( \gamma + q + 1 )   \beta _5  = 0   .
\end{equation}
We get a single additional  variational symmetry
\begin{multline}
Z_{4}
=  { \gamma+ q + 1 \over 2 } X_{4}  +   { \gamma - 3 \over 2 } X_{5}
\\
=(3\gamma + 2 q - 1 )  t\frac{\partial}{\partial t}
+(\gamma-3)s\frac{\partial}{\partial s}
+(\gamma+1+q)\varphi\frac{\partial}{\partial\varphi}
\end{multline}
in addition to the symmetries  (\ref{Noether_symmetry_0_1})
and  (\ref{Noether_symmetry_0_2}).

 Using Noether's theorem, we compute the components of the conservation law
\begin{multline}
T_{4}^{t}
=
   ( 3 \gamma   + 2 q   - 1 )  t \left(
 { \varphi_{t}^{2}   \over 2 }
+ { A_{0}  s^q  \over \gamma -1}  \varphi_{s}^{1-\gamma}
\right)
+    (  \gamma - 3 )    s \varphi _{t} \varphi_{s}
-       (\gamma  + q + 1)      \varphi\varphi_{t}     ,
\\
T_{4}^{s}
=
  ( 3 \gamma + 2 q - 1 ) t  A_{0}  s^q \varphi_{t}   \varphi_{s}^{-\gamma}
\\
+   (  \gamma - 3 )  s \left(
 -  {   \varphi_{t}^{2} \over 2 }
+ {  \gamma    A_{0} s^q    \over \gamma -1}     \varphi_{s}^{1-\gamma}
\right)
-  (\gamma  + q + 1 )  A_{0}  s^q \varphi     \varphi_{s}^{-\gamma}      .
\end{multline}
In the gas dynamics variables we get
\begin{multline}
T_{4}^{t}
=
   ( 3 \gamma   + 2 q   - 1 )  t \left(
 { u ^{2}   \over 2 }
+ { A_{0}  s^q  \over \gamma -1}  \rho ^{\gamma - 1 }
\right)
+    (  \gamma - 3 )    s    { u \over \rho  }
-       (\gamma  + q + 1)      \varphi  u   ,
\\
T_{4}^{s}
=
  ( 3 \gamma + 2 q - 1 ) t  A_{0}  s^q    u    \rho ^{ \gamma}
\\
+   (  \gamma - 3 )  s \left(
  -  {   u ^{2} \over 2 }
+  {  \gamma    A_{0} s^q    \over \gamma -1}     \rho ^{\gamma -1 }
\right)
-  (\gamma  + q + 1 )  A_{0}  s^q \varphi     \rho ^{\gamma}     .
\end{multline}

To rewrite this conservation law in Eulerian coordinates  
we use the transformation rule  (\ref{transformation_rule})
and
\begin{equation}
s = q \rho { S \over S_x}   .
\end{equation}
We obtain  
\begin{multline*}
 {}^{e}T_{4}^{t}
=  (3\gamma+2q-1) t  \left(
 { \rho u ^{2}   \over 2 }
+ { S \over \gamma -1}  \rho ^{\gamma}
\right)
+ (\gamma-3) q \rho u  { S \over S_x}
 -   ( \gamma + q  + 1 )  x \rho u   ,
\\
 {}^{e}T_{4}^{x}
=  (3\gamma+2q-1) t  \left(
 { \rho u ^{2}   \over 2 }
+ {  \gamma  S \over \gamma -1}  \rho ^{\gamma}
\right) u
\\
+ (\gamma-3) q \rho   { S \over S_x}
 \left(
 { u ^{2}   \over 2 }
+ {  \gamma  S \over \gamma -1}  \rho ^{\gamma-1}
\right)
 -   ( \gamma + q  + 1 )   x  ( \rho u ^2+ S  \rho ^{\gamma} )     .
\end{multline*}



\subsubsection{Case $\gamma=3$}

For symmetries (\ref{kernel1}),  (\ref{kernel2}) and      (\ref{additional_2_0})
substitution of  the generator
\begin{equation}
X = \sum  _{i=1} ^5   \beta _i  X_i      + \beta _*  X_*
\end{equation}
into (\ref{eq:Noether})
gives the condition
\begin{equation}
  (  q + 4 )   \beta _5  = 0
\end{equation}
for variational and divergent symmetries. There are two possibilities.


\medskip

\noindent
a) For $q\neq-4$ there is no additional generators.
We get the conservation laws which were described in points \ref{point_kernel_1} and  \ref{point_kernel_2}.


\medskip

\noindent
b) If $q=-4$, there is one additional  variational symmetry
\begin{equation}
Z_{4}  = { 1 \over 2 } X_5 = t\frac{\partial}{\partial t}+s\frac{\partial}{\partial s}  .
\end{equation}
The corresponding conservation law has densities
\begin{multline}
T_{4}^{t}
=    t \left(
{  \varphi_{t}^{2}   \over 2 }
+  { A_{0}    s^{q}       \over  \gamma - 1  }  \varphi_{s}^{ 1 - \gamma }
\right)
+   s \varphi_{t} \varphi_{s}   ,
\\
T_{4}^{s}
=   t   A_{0}  s^{q}    \varphi_{t}   \varphi_{s}^{- \gamma}
+ s \left(
-    { \varphi_{t}^{2}   \over 2 }
+   {  \gamma    A_{0}     s^{q}   \over   \gamma - 1   }   \varphi_{s}^{1 - \gamma}
\right) .
\end{multline}
In the gas dynamics variables we get
\begin{multline}
T_{4}^{t}
=    t \left(
{  u ^{2}   \over 2 }
+  { A_{0}   s^{q}         \over   \gamma - 1   }  \rho  ^{ \gamma - 1 }
\right)
+  s    {  u   \over \rho }  ,
\\
T_{4}^{s}
=   t   A_{0}  s^{q}    u    \rho ^{\gamma}
+ s \left(
-    {   u  ^{2}   \over 2 }
+   {    \gamma   A_{0}     s^{q}   \over  \gamma - 1    }   \rho ^{\gamma -1}
\right) .
\end{multline}
These components get transformed into Eulerian coordinates as 
\begin{multline*}
{}^{e}T_{4}^{t}
=   t  \left(
 { \rho u ^{2}   \over 2 }
+ { S \over  \gamma -1  }  \rho ^{\gamma}
\right)
 + q  \rho u    { S \over S_x} ,
\\
{}^{e}T_{4}^{x}
=   \left(
t \rho u   +  q \rho     { S \over S_x}
\right)
\left(
 { u ^{2}   \over 2 }   + {  \gamma  S \over  \gamma - 1  }  \rho ^{\gamma -1}
\right)    .
\end{multline*}


\subsection{Entropy $ S (s) =A_{0}e^{q s} $}

This  assumption for the entropy
is given  in Eulerian coordinates  by the differential constraint
\begin{equation}
S_{x}=\rho q S.
\end{equation}
It is possible to check 
that the overdetermined system of the gas dynamics equations 
and the latter constraint is involutive.

The classifying equations  (\ref{conditions})  lead  to
\begin{subequations}
\begin{gather}
    (\gamma+1)    k_{1}  -  2k_{3}   -   q k_{8}   =  0 ,     \\
k_{7} = 0  , \\
 (\gamma-3)   k_{2}   =  0  .
\end{gather}
\end{subequations}
For both cases  $\gamma=3$ and $\gamma\neq3$
we find one additional symmetry:
\begin{equation}        \label{additional_3_0}
X_{5}
=q Y_{3}-2Y_{8}
=qt\frac{\partial}{\partial t}
-2\frac{\partial}{\partial s} .
\end{equation}


\subsubsection{Case $\gamma\protect\neq3$}

Substitution of the generator
\begin{equation}
X = \sum  _{i=1} ^5   \beta _i  X_i
\end{equation}
into (\ref{eq:Noether})
gives that the condition for Noether symmetries as
\begin{equation}
 ( 3 - \gamma )    \beta _4
 - q     \beta _5  = 0   .
\end{equation}
There is the variational symmetry
\begin{equation}
Z_{4}
=  {  q  \over 2 } X_{4}  +   { 3 - \gamma   \over 2 } X_{5}
=  2qt\frac{\partial}{\partial t}
+(\gamma-3)\frac{\partial}{\partial s}
+q\varphi\frac{\partial}{\partial\varphi}
\end{equation}
in addition to the symmetries  (\ref{Noether_symmetry_0_1})
and  (\ref{Noether_symmetry_0_2}),
which gives the conserved vector
\begin{multline}
T_{3}^{t}
=
   2 q    t \left(
 { \varphi_{t}^{2}   \over 2 }
+ { A_{0}  e^{qs}   \over \gamma -1}  \varphi_{s}^{1-\gamma}
\right)
-  q     \varphi \varphi_{t}
+     (  \gamma - 3 )      \varphi_{t} \varphi_{s}     ,
\\
T_{3}^{s}
=
   2 q  t  A_{0}    e^{qs}         \varphi_{s}^{-\gamma}    \varphi_{t}
-  q   A_{0} e^{qs}    \varphi   \varphi_{s}^{-\gamma}
+    (  \gamma - 3 )  \left(
 -  {  \varphi_{t}^{2} \over 2 }
+   {  \gamma    A_{0}   e^{qs}    \over \gamma -1}      \varphi_{s}^{1-\gamma}
\right)   .
\end{multline}
It can be rewritten in the gas dynamics variables as
\begin{multline}
T_{3}^{t}
=   2 q    t \left(
 { u  ^{2}   \over 2 }
+ { A_{0}  e^{qs}   \over \gamma -1}  \rho ^{\gamma-1}
\right)
-  q     \varphi u
+     (  \gamma - 3 )      { u \over \rho }    ,
\\
T_{3}^{s}
=
   2 q  t  A_{0}    e^{qs}        \rho ^{\gamma}  u
-  q   A_{0} e^{qs}    \varphi   \rho ^{\gamma}
+    (  \gamma - 3 )  \left(
 -  {  u ^{2} \over 2 }
+   {  \gamma   A_{0}   e^{qs}     \over \gamma -1}      \rho ^{\gamma-1}
\right)     .
\end{multline}
We can also give this conservation law in Eulerian coordinates
\begin{multline*}
{}^{e}T_{3}^{t}
=  2 q    t \left(
 { \rho u  ^{2}   \over 2 }
+ { A_{0}  e^{qs}   \over \gamma -1}  \rho ^{\gamma}
\right)
-  q    x  \rho   u
+     (  \gamma - 3 )      { u }    ,
\\
{}^{e}T_{3}^{x}
=  2 q    t \left(
 { \rho    u  ^{2}   \over 2 }
+ { \gamma A_{0}  e^{qs}   \over \gamma -1}  \rho ^{\gamma}
\right)  u
-  q     x   (     \rho u^2  +       A_{0} e^{qs}      \rho ^{\gamma}    )
 +    (  \gamma - 3 )  \left(
   {  u ^{2} \over 2 }
+   {  \gamma     A_{0}   e^{qs}     \over \gamma -1}    \rho ^{\gamma-1}
\right)    .
\end{multline*}



\subsubsection{Case $\gamma=3$}

In this case we consider symmetries  (\ref{kernel1}),  (\ref{kernel2}) and      (\ref{additional_3_0}).
Looking for  Noether  symmetries, 
we substitute 
\begin{equation}
X = \sum  _{i=1} ^5   \beta _i  X_i        + \beta _*  X_*
\end{equation}
into (\ref{eq:Noether}) 
and obtain
gives
\begin{equation}
   q     \beta _5  = 0   .
\end{equation}
There are no additional Noether symmetries. 
Therefore  we do not any conservation laws in addition to those 
given in points  \ref{point_kernel_1} and   \ref{point_kernel_2}.

\subsection{Discussion}

First of all we note that to select variational or divergent symmetries 
using property (\ref{eq:Noether}) of an admitted symmetry
it was important to consider a generator $X$ in its general form, 
i.e. as a linear combination of all admitted symmetries.

Complete Lie group classification of the gas dynamics equation  
in  the Lagrangian coordinates   (\ref{eq:one-D-gas_Lagrange})
allows us to find all  conservation laws which can be found using Noether's theorem and admitted symmetries.
The group classification has three cases of the entropy
for which there exist additional symmetries.
In Eulerian coordinates these three cases are defined by  differential constraints of first or second order.
Notice that the overdetermined systems which consist of the gas dynamics equations 
and one of the considered differential constraints are involutive.
The authors of \cite{bk:WebbZank[2009],bk:Webb2018} also found conservation laws
corresponding to special forms of the entropy.
Here the symmetry nature of these conservation laws is explained.

In contrast to \cite{bk:SjobergMahomed2004}
the conservation laws, obtained in this paper,  are local. It should be also noted that this
conservation laws are naturally derived: 
their counterparts in Lagrangian coordinates were derived directly 
using Noether's theorem without any additional assumptions.
We should also mention that in contrast to two-dimensional Lagrangian gas dynamics
the special cases of the entropy in Lagrangian coordinates are given explicitly.
In the two-dimensional case \cite{Kaptsov_press} the entropy is arbitrary,
but the admitted symmetry operators  contain  functions 
satisfying quasilinear partial differential equations.

\section{Finite-difference models}

\label{Finite_difference}

\subsection{Invariance and Euler coordinates}

The first problem in discretization of differential equations is the
choice of difference mesh. For discretization of the gas dynamics
system (\ref{Euler_GD_rho}),(\ref{Euler_GD_u}),(\ref{Euler_GD_p}),
which is given in Euler coordinates, the simplest choice seems to be
an orthogonal mesh in $ (t,x) $ plane. 
However, this mesh is not invariant 
that destroys invariance of difference equations considered on such mesh. 
Indeed, as it was shown in   \cite{Dorodnitsyn1989,  Dorodnitsyn2011} 
the necessary condition for a mesh to preserve its invariance under a group transformation 
generated by the operator
\begin{equation}
X=
\xi ^t  {\partial \over \partial t }
+  \xi ^x  {\partial \over \partial x }
+ ...
\end{equation}
is the following:
\begin{equation}  \label{orthogonality}
D_{+h} (\xi^t) = -D_{+\tau} (\xi^x) ,
\end{equation}
where $D_{+h}$ and $D_{+\tau}$ are the operators of difference
differentiation in $x$ and $t$ directions respectively.

The system   (\ref{Euler_GD_rho}),(\ref{Euler_GD_u}),(\ref{Euler_GD_p})   admits
\cite{bk:Ovsiannikov1978} the $6$-parameter Lie symmetry group of point
transformations  that corresponds to the following Lie algebra of
infinitesimal operators:
\begin{multline}     \label{Euler_symmetries}
X_1 =\frac{\partial}{\partial t}  ,
\qquad
X_2 =\frac{\partial}{\partial x} ,
\qquad
X_3 =t\frac{\partial}{\partial t} + x\frac{\partial}{\partial x} ,
\qquad
X_4 =t\frac{\partial}{\partial x}+ \frac{\partial}{\partial u}   ,
\\
X_5 =x\frac{\partial}{\partial x} + u\frac{\partial}{\partial u} -
2\rho\frac{\partial}{\partial {\rho}}  ,
\qquad
X_6 = \rho \frac{\partial}{\partial {\rho}}   + p \frac{\partial}{\partial p}  .
\end{multline}
In the special case $\gamma = 3$ there  is one more symmetry 
\begin{equation}         \label{Euler_symmetry_additional}
X_7 
=t^2\frac{\partial}{\partial t}+ tx\frac{\partial}{\partial x}
+ (x-tu)\frac{\partial}{\partial u}-t\rho\frac{\partial}{\partial
{\rho}}- 3tp\frac{\partial}{\partial p} . 
\end{equation}

It is
easy to see, the  Galileo transformation, given by the operator $X_4$, 
does not satisfy the  criterion (\ref{orthogonality}).
The same is true for $X_7$.
It means that one  should look for an invariant moving mesh in Eulerian   coordinates.

To obtain an invariant moving mesh we chose   the following  difference stencil with two time layers:

\begin{itemize}

\item

independent variables:
$$
 t = t_j  ,
\quad
 \hat{t}=   t _{j+1} ;
\quad
 x = x_{i} ^{j}  ,
\quad
x_+   = x_{i+1} ^{j} ,
\quad
 \hat{x} = x_{i} ^{j+1},
\quad
 \hat{x}_+ = x_{i+1} ^{j+1};
$$

\item

dependent variables in the nodes of the mesh (as $x$):
$$
  u   ,
\quad
u_+    ,
\quad
 \hat{u} ,
\quad
 \hat{u}_+ ;
\quad
  \rho      ,
\quad
 \rho   _-    ,
\quad
 \hat{\rho} ,
\quad
 \hat{\rho}_-   ;
 \quad
p    ,
\quad
p _-  ,
\quad
\hat{p} ,
\quad
 \hat{p}_-   . 
$$
\end{itemize}
Then, we find the finite-difference invariants for symmetries (\ref{Euler_symmetries}) 
as the solutions of system of linear equations
\begin{equation}
X _i   I  (   t   ,    \hat{t}  ,  
 x  ,   x_+    ,   \hat{x}  , \hat{x}_+ , 
 ...  , 
 \rho      ,  \rho   _-    , \hat{p} ,  \hat{p}_-    )  = 0 ,    
\qquad   
i= 1, . . . , 6 . 
\end{equation}
Here we assume that the operator is prolonged on all variables of the stencil. 
There are 12 functionally independent invariants
\begin{equation*}
\frac{\hat{h}_+}{h_+}  ,
\qquad
\frac{\tau}{h_+}   \sqrt{\frac{p}{\rho}}  ,
\qquad
\sqrt{\frac{\rho}{p}}  \left( \frac{ \hat{x} - x } { \tau } - u \right)  ,
\end{equation*}
\begin{equation}
\sqrt{\frac{\rho}{p}}(u_+ -u) ,
\qquad
\sqrt{\frac{\rho}{p}}(\hat{u}-u) ,
\qquad
\sqrt{\frac{\rho}{p}}(\hat{u}_+ - \hat{u}) ,
\end{equation}
\begin{equation*}
\frac{p_+}{p}  ,
\qquad
\frac{\hat{p}}{p}  ,
\qquad
\frac{\hat{p}_+}{\hat{p}}  ,
\qquad
\frac{\hat{\rho}}{\rho}  ,
\qquad
\frac{\hat{\rho}_+}{\hat{\rho}}  ,
\quad
\frac{{\rho_+}}{\hat{\rho}}  ,
\end{equation*}
where  $ \tau =  \hat{t}   -   t   $,
$ h_+   =  x _ +    -  x   $
and
$ \hat{h} _+   =   \hat{x}  _ +    -  \hat{x}   $.

These invariants suggest,  for example,
 an  invariant moving mesh given by
\begin{equation}
\sqrt{\frac{\rho}{p}}  \left( \frac{ \hat{x} - x } { \tau } - u  \right) = 0
\end{equation}
or,  equivalently,
\begin{equation}
\frac{ \hat{x} - x } { \tau } = u .
\end{equation}
In the continuous limit it corresponds to the evolution of the spacial  variable $x$ given as
\begin{equation}
\frac{dx}{dt} =u  .
\end{equation}
Thus, we arrive at choosing the mass Lagrangian coordinates.

Below we introduce a  scheme, which is invariant with respect to 
symmetries of 1+1 D gas dynamics in the case  $ \gamma  \neq  3$.

\subsection{Popov-Samarskii scheme}

One of the best known numerical schemes for one-dimensional gas
dynamics in the  mass mass coordinates is  {\it completely conservative}  scheme  
introduced by Yu.~P.~Popov and A.~A.~Samarskii \cite{Popov1969, Samarskii1975}.
It was derived as a scheme which preserves
conservation laws of mass, momentum and energy. Usually it is given
for gas dynamics equations in the form
\begin{subequations}            \label{gas_dynamics_ for_SP}
\begin{gather}
\left(  { 1 \over \rho }  \right) _t = u   _s     ,
\label{equation_rho}
\\
u _t  +     p_{s}    = 0    ,
 \label{equation_u}
\\
\varepsilon  _t =   -  p      u  _s    ,
 \label{equation_energy}
\\
x_t =    u    , 
 \label{equation_coordinate}
\end{gather}
\end{subequations}
where 
\begin{equation}       \label{Lagrange_state}
\varepsilon    =   {  p   \over ( \gamma -1) \rho }  ,
\qquad
\gamma   \neq 0, 1
\end{equation}
is the internal energy of the gas. 
For the polytropic gas 
equation   (\ref{equation_energy}) is equivalent  to (\ref{Lagrange_GD_S})  and (\ref{Lagrange_GD_p}).
However, the scheme can also   be used for other equations of  the internal energy
$
\varepsilon    = \varepsilon   ( \rho ,  p ) .
$

The gas dynamics system is presented with  the help of the internal energy
$ \varepsilon    $
to emphasize importance of the relation
\begin{equation}     \label{work}
\varepsilon  _t =   -  p   \left(  { 1 \over \rho }  \right) _t
\end{equation}
for qualitatively correct discretization.
This relation shows that the change of the internal energy happens
due to the work done by the pressure forces.

\subsubsection{Notations}

\label{scheme_notations}

Let us introduce notations which are needed to present the scheme. 
In  the coordinates $(t,s)$ we can consider an orthogonal mesh. 
For simplicity we chose a uniform mesh for the Lagrange mass variable $s$, i.e.
neighbouring spatial step
\begin{equation}
h^s = s_{i+1} - s _i
\qquad
\mbox{and}
\qquad
h^s _- = s_{i} - s _{i-1}
\end{equation}
will be equal:   $  h^s  = h^s  _- $.
Of course, it is also possible to consider   nonuniform  steps
$
h^s  _i  = s_{i+1} - s _i  .
$

Variables $u  $ and $x$  are taken in the nodes of the mesh as
\begin{equation}
u =  u_ i  ^j ,
\qquad
u_+ =  u_{i+1} ^j  ,
\qquad
\hat{u}  =  u_ i  ^{j+1}  ,
\qquad
\hat{u} _+ =  u_{i+1} ^{j+1}    .
\end{equation}
The time derivative and forward spatial derivatives will be
\begin{equation}
u_t = { \hat{u}   -  u   \over  \tau  } ,
\qquad
u_s = { u_{i+1}  ^j   - u _i ^j    \over  s_{i+1} - s_i   }
=  { u_{+}   - u   \over   h ^s  }  .
\end{equation}


Variables $\rho  $, $p $  and $ \varepsilon $
are assigned to the midpoints $i-1/2 $,  $ i+1/2 $, $ i+3/2 $ and so on.
For example,
\begin{equation}
\rho_-   =  \rho  _ {i-1/2}  ^j  ,
\qquad
\rho   =  \rho  _ {i+1/2}  ^j  ,
\qquad
\rho  _+ =  \rho  _ {i+3/2} ^j  .
\end{equation}
For their backward spatial derivatives we take
\begin{equation}
 p_{\bar{s}} = { p _{ i+1/2 }  ^j -   p _{ i-1/2 }  ^j \over  h^s     } ,
\end{equation}
For pressure $p$ we will need the linear interpolation value for the nodes of the mesh.
For example,  for the node $ ( t_j ,   s_i ) $ we have
\begin{equation}
  p_{*}    = p_{*i}  ^j   = {   p _{ i-1/2 }  ^j   +    p _{ i+1/2 }  ^j   \over   2     }
  .
\end{equation}

Weighted values for all variables  will be denoted as
\begin{equation}
 y    ^{(\alpha)}    =  \alpha   \hat{y} + ( 1 - \alpha ) y   ,
\qquad
0 \leq \alpha \leq 1   .
\end{equation}

\subsubsection{The scheme  and its properties}

\label{scheme_system}

Popov-Samaskii   scheme   \cite{Popov1969}
\begin{subequations}      \label{Samarskii_Popov_scheme}
\begin{gather}
\left(  { 1 \over \rho }  \right) _t = ( u  ^{(0.5)}   )_s
   \label{equation_discrete_mass}
\\
u _t =   -     p_{\bar{s}}    ^{(\alpha)}   ,
\label{equation_discrete_velocity}
\\
\varepsilon  _t =   -  p ^{(\alpha)} (    u  ^{(0.5)} ) _s    ,
  \label{equation_discrete_energy}
\\
x_t =    u  ^{(0.5)}     .
 \label{equation_discrete_coordinate}
\end{gather}
\end{subequations}
is invariant with respect to symmetries   (\ref{Euler_symmetries}) extended to varaible $s$:
\begin{multline}        \label{Lagrange_symmetry}
X_1 =\frac{\partial}{\partial t}   ,
\qquad
X_2 =\frac{\partial}{\partial x}  ,
\qquad
X_3 =t\frac{\partial}{\partial t} + s\frac{\partial}{\partial s}   + x
\frac{\partial}{\partial x}   ,
\qquad
    X_4 =   t \frac{\partial}{\partial x}  +    \frac{\partial}{\partial u} ,
\\
X_5 = - s\frac{\partial}{\partial s} +  x\frac{\partial}{\partial x}
+ u\frac{\partial}{\partial u} - 2\rho\frac{\partial}{\partial {\rho}}  ,
\qquad
X_6 =  s \frac{\partial}{\partial s}
+ \rho\frac{\partial}{\partial {\rho}} +  p\frac{\partial}{\partial p}  .
\end{multline}
It also admit translations  of the new dependent variable $s$
\begin{equation}      \label{Lagrange_symmetry_s}
X_{7} = \frac{\partial}{\partial s}  .
\end{equation}

However, it is not invariant for the projective operator
\begin{equation}        \label{Lagrange_symmetry_additional}
X_8
= t^2 \frac{\partial}{\partial t}
+ (x-tu) \frac{\partial}{\partial u}
- t \rho \frac{\partial}{\partial {\rho}}
- 3 t p \frac{\partial}{\partial p}
+ t x  \frac{\partial}{\partial x}   ,
\end{equation}
which is admitted by   (\ref{gas_dynamics_ for_SP}),     (\ref{Lagrange_state})
 in the special case  $\gamma = 3$.


\medskip

Let us discuss the properties of this scheme.
In the general case   $ \varepsilon   ( \rho ,  p )$,
i.e. not only for the ideal gas (\ref{Lagrange_state}),
the scheme (\ref{Samarskii_Popov_scheme}) has the following   conservation laws:

\begin{enumerate}

\item
Conservation of mass (\ref{equation_discrete_mass}),
which  manifests the consistency  of the relations
\begin{equation}
 { h_i  ^s \over \rho _{ i +1/2} ^j }  =   {  x _{i+1} ^j    -   x_i  ^j     }  ,
\qquad
 { h_i  ^s \over \rho _{ i +1/2} ^{j+1} }  =   {  x _{i+1} ^{j+1}    -   x_i  ^{j +1}    }
\end{equation}
with the evolution equation       (\ref{equation_discrete_coordinate}).

\item

Conservation of  momentum
\begin{equation}
\left[
    u
 \right]    _t
+
  [      p  ^{(\alpha)}    ]  _{\bar{s}}
=   0  .
\end{equation}

\item

Conservation of  energy
\begin{equation}
\left[ \varepsilon   + {  u ^2  +  u_+ ^2    \over 4 }       \right]   _t
+
   [ p _{*} ^{(\alpha)}    u ^{(0.5)}  ] _s
=   0   .
\end{equation}

\item
The law for the  center  of mass motion
\begin{equation}
\left[
x  -   t   u
 \right]    _t
-
  [    t   ^{(0.5)}    p  ^{(\alpha)}    ]  _{\bar{s}}
=   0  ,
\end{equation}
 which was not mentioned by the authors \cite{Popov1969}.

\end{enumerate}

These discrete conservation laws correspond to continuous conservation laws for 

\begin{enumerate}

\item

mass
 \begin{equation}       \label{Lagrange_mass}
     \left[ \frac{1}{\rho} \right]_t  -    \left[   u  \right] _s = 0   ;
\end{equation}

\item

 momentum
\begin{equation}      \label{Lagrange_momentum }
     \left[ u   \right]_t + \left[  p  \right] _s = 0   ;
\end{equation}

\item

 energy

\begin{equation}     \label{Lagrange_energy}
     \left[   \frac{u^2}{2} +  \frac{p}{(\gamma -1)\rho} \right] _t
+  \left[  p u \right] _s = 0    ;
\end{equation}

\item

center of mass motion
\begin{equation}      \label{Lagrange_center_mass}
     [  x - t u ] _t - [ t p] _s = 0    .
\end{equation}

\end{enumerate}

For the polytropic gas (\ref{Lagrange_state}) with $\gamma = 3$
there exist two additional conservation laws



\begin{equation}      \label{Lagrange_additional_1}
     \left[
2t \left(\frac{u^2}{2} +  \frac{p}{(\gamma -1)\rho} \right) - xu
\right] _t +  \left[ 2tpu - xp \right] _s = 0  ,
\end{equation}


\begin{equation}     \label{Lagrange_additional_2}
     \left[
t^2 \left(\frac{u^2}{2} +  \frac{p}{(\gamma -1)\rho} \right) - txu +
\frac{x^2}{2} \right]_t +  \left[ t^2 pu - txp \right]_s = 0  ,
\end{equation}
which  the scheme (\ref{Samarskii_Popov_scheme}) does not have.

The scheme does not preserve entropy, or 
\begin{equation}       \label{property_nonlinear}
S = {  {p}  \over   {\rho}  ^{\gamma} }   =  \mbox{const}   
\end{equation}
for the streamlines of the flow.
However, it holds the relation
\begin{equation}
{ \Delta   p  \over   p ^{(\alpha)}     }
= \gamma
{ \Delta   {\rho}    \over   \rho   ^{(\alpha)}      }  ,
\end{equation}
where   $  \Delta   p  = \hat{p} - p $  and  $  \Delta   \rho   = \hat{\rho} - \rho  $. 
This discrete relation approximates its differential counterpart 
\begin{equation}
{ p _t    \over   p    }
= \gamma
{   {\rho}  _t   \over   \rho   }  ,
\end{equation}
with the order $0(\tau)$. Thus, we get an error for the entropy evolution.


It should also be noted that the scheme   (\ref{Samarskii_Popov_scheme})
 has the relation
\begin{equation}     \label{discrete_work}
\varepsilon  _t =   -  p ^{(\alpha)}   \left(  { 1 \over \rho }  \right) _t  ,
\end{equation}
which is a  qualitatively correct discretization of the  continuous
case relation   (\ref{work}).
It represents  a correct balance of the internal energy.

\subsection{Invariance of finite-difference schemes}

Let us show how one can use invariant to construct finite-difference schemes.
For the case ${\gamma  \neq   3 }$  we will show how the scheme, given in the previous point,
can be expressed in term of the invariants.
For the case ${\gamma  =   3} $ we will suggest an invariant scheme.

\subsubsection{Case  ${\gamma  \neq   3 }$ }

\label{scheme_general_gamma}

We consider an orthogonal mesh in $ (t,s) $ coordinates
and specify the stencil variables as

\begin{itemize}

\item

independent variables:
$$
t = t_j  ,
 \quad
\hat{t} = t _{j+1} ,
 \quad
s = s_i  ,
\quad
s_+  =   s_{i+1}  ,
\quad
 s_-  =    s_{i-1};
$$

\item

dependent variables in the nodes of the mesh:
$$
  u = u_{i} ^{j}  ,
\quad
u_+   = u_{i+1} ^{j} ,
\quad
 \hat{u} = u_{i} ^{j+1},
\quad
 \hat{u}_+ = u_{i+1} ^{j+1} ,
 \quad
 x   ,
\quad
x_+  ,
\quad
 \hat{x} ,
\quad
 \hat{x}_+ ;
 $$

\item

dependent variables in the midpoints of the mesh:
$$
  \rho    =  \rho   _{i} ^{j}  ,
\quad
 \rho   _-   =  \rho   _{i-1} ^{j} ,
\quad
 \hat{\rho} =  \rho   _{i} ^{j+1},
\quad
 \hat{\rho}_- =  \rho   _{i-1} ^{j+1} ,
 \quad
p    ,
\quad
p _-  ,
\quad
\hat{p} ,
\quad
 \hat{p}_-    .
$$
\end{itemize}


In the space of 21 stencil variables there are 14 invariants for
7 symmetries  (\ref{Lagrange_symmetry}),  (\ref{Lagrange_symmetry_s}):
\begin{equation*}
I_{1} = { h^s _-   \over h^s }    ,
 \quad
I_{2} = \frac{\tau}{h^s}  \sqrt{ {\rho}{p} } ,
 \quad
I_{3} = \sqrt{ \frac{\rho}{p} } \left( \frac{ \hat{x} - x } { \tau } - u  \right),
\quad
I_{4} = \sqrt{ \frac{{\rho}}{{p}} }   \left( \frac{ \hat{x} - x } { \tau } - \hat{u} \right),
\end{equation*}
\begin{equation*}
I_{5} = \sqrt{ \frac{\rho}{p} }(u_+ -u),
\quad
I_{6} = \sqrt{\frac{\rho}{p}}(\hat{u}_+ - \hat{u}),
\quad
I_{7} = {  \rho ( x_+ - x)   \over h^s }   ,
\quad
I_{8} = {  \hat{\rho} ( \hat{x}_+ - \hat{x} )   \over h^s }   ,
\end{equation*}
\begin{equation*}
I_{9} = \frac{\rho_-}{\rho}  ,
\quad
I_{10} = \frac{\hat{\rho}}{\rho}  ,
\quad
I_{11} = \frac{\hat{\rho}_-}{\hat{\rho}}  ,
\quad
I_{12} = \frac{p_-}{p}  ,
\quad
I_{13} = \frac{\hat{p}}{p}  ,
\quad
I_{14} = \frac{\hat{p}_-}{\hat{p}}    .
\end{equation*}

One can find the scheme
(\ref{Samarskii_Popov_scheme}) approximating the gas dynamics system
(\ref{gas_dynamics_ for_SP}) with the help of these invariants as
\begin{equation*}
{ 1 \over I_{10} }  - 1 =   I_{2}    {   I_{5} +  I_{6}   \over 2 }   ,
\end{equation*}
\begin{equation*}
I_{3}  -  I_{4}  =  - I_{2}   \left(
\alpha \left(   I_{13}  -    { I_{13}  I_{14} }      \right)
+ ( 1 - \alpha ) ( 1  - I_{12} )
\right)   ,
\end{equation*}
\begin{equation*}
{ 1 \over  \gamma -1 } \left( {  I_{13}  \over  I_{10}  }  - 1 \right)
= -   I_{2}    (     \alpha   I_{13}    +   ( 1 - \alpha ))     {   I_{5} +  I_{6}   \over 2 }    ,
\end{equation*}
\begin{equation*}
 I_{3} +  I_{4}   = 0 .
\end{equation*}

\subsubsection{Case  ${\gamma =  3 }$ }

For the case ${\gamma = 3 }$ we can suggest invariant schemes,
but have difficulties to  find invariant schemes  with conservation laws.

In comparison to the previous point \ref{scheme_general_gamma}
we have one more symmetry, namely   (\ref{Lagrange_symmetry_additional}).
Therefore, we get one invariant less.

There obtain  13 invariants
\begin{equation*}
J_{1} = { h^s _-   \over h^s }    ,
 \quad
J_{2} = \frac{\tau}{h^s}   ( {\rho}{p} \hat{\rho} \hat{p}  )   ^{1 \over 4}  ,
 \quad
J_{3} = \sqrt{ \frac{\rho}{p} } \left( \frac{ \hat{x} - x } { \tau } - u  \right),
\quad
J_{4} = \sqrt{ \frac{\hat{\rho}}{\hat{p}} }   \left( \frac{ \hat{x} - x } { \tau } - \hat{u} \right),
\end{equation*}
\begin{equation*}
J_{5} = \sqrt{ \frac{\rho}{p} }
\left( { \frac{h_+}{\tau} } +  u_+ -u \right),
\quad
J_{6} = \sqrt{ \frac{\hat{\rho}}{\hat{p}} }
\left( - { \frac{\hat{h}_+}{\tau} } +  \hat{u}_+ -\hat{u} \right),
\quad
J_{7} = {  \rho ( x_+ - x)   \over h^s }   ,
\end{equation*}
\begin{equation*}
J_{8} = {  \hat{\rho} ( \hat{x}_+ - \hat{x} )   \over h^s }   ,
\quad
J_{9} = \frac{\hat{p}}{p}   \left(  \frac{\rho}{\hat{\rho}} \right)^3    ,
\quad
J_{10} = \frac{\rho_-}{\rho}  ,
\quad
J_{11} = \frac{\hat{\rho}_-}{\hat{\rho}}  ,
\quad
J_{12} = \frac{p_-}{p}  ,
\quad
J_{13} = \frac{\hat{p}_-}{\hat{p}}    .
\end{equation*}

There are many possibilities to approximate the gas dynamics system
  (\ref{gas_dynamics_ for_SP}),   (\ref{Lagrange_state})
 with the help of these invariants.
Let us suggest the following explicit invariant scheme
\begin{subequations}   \label{explicit_scheme}
\begin{gather}
   \hat{\rho}  (  \hat{x} _+   - \hat{x} )  =  {\rho}   (   {x} _+   - x )     ,
 \label{explicit_scheme_rho}
\\
{ \frac{\hat{u} - u}{\tau}  } =
-
\left(
{ \frac{\hat{\rho}}{\rho}}
 \right)  ^2
{ \frac{p   - p_- }{h ^s} }     ,
\label{explicit_scheme_u}
\\
 \frac{\hat{p}}{\hat{\rho} ^3} =  \frac{p}{{\rho} ^3}
\label{explicit_scheme_p}  ,
\\
 \frac{ \hat{x} - x }{ \tau } =  u
\label{explicit_scheme_x}  .
\end{gather}
\end{subequations}
In term of the invariants this scheme is written as
\begin{equation*}
J_{7} =    J_{8}      ,
\end{equation*}
\begin{equation*}
 J_{4}  =     J_{2}    J_{9} ^{-3/4}      (    1 - J_{12}  )   ,
\end{equation*}
\begin{equation*}
J_{9} = 1   ,
\end{equation*}
\begin{equation*}
J_{3} = 0     .
\end{equation*}

The scheme keeps an entropy along streamlines
(\ref{property_nonlinear}) and possesses conservation of mass
(\ref{explicit_scheme_rho}).
Note that the first equation can be rewritten as
\begin{equation}
  {1\over \tau  } \left( {  1 \over \hat{\rho}  }   -     {  1 \over {\rho}  }  \right)
 =  { u _+ - u   \over h^s }     .
\end{equation}
Let us note that implicit invariant schemes are  also possible.

\section{Concluding remarks}

\label{Concluding_remarks}

In the present paper there were considered 
Lie point symmetries of the one-dimensional gas dynamics equations of  the polytropic   gas 
 in Lagrangian coordinates. Complete Lie group
classification of these equations reduced to the scalar second-order
PDE  is performed. The
classification parameter is the entropy. It was shown that for
the isentropic gas and for the  entropy depending of mass coordinate as power
and exponential law there exist extensions of the symmetry  properties. 
Therefore, there are additional  conservation laws.
 For these cases Noether's theorem was applied 
for constructing  conservation laws for the  scalar
second-order PDE. The conservation laws  were later represented in the gas
dynamics variables.

For the basic adiabatic   case invariant and conservative difference
schemes were  discussed. The whole set of difference invariants was
constructed for the  general case and for the extension case $\gamma=3$. 
It was shown  that the Popov-Samarskii scheme is invariant for the basic case symmetries, 
but does not allow the additional symmetry which exists for $\gamma=3$.
This scheme possesses the whole set of conservation laws, excluding
two additional laws which hold for the  exceptional value of $\gamma$. It also does not
preserve the entropy along streamlines. The scheme which preserve all
symmetries for $\gamma =3 $ was developed. This scheme
holds exactly the entropy along the streamlines but has only one conservation law, 
namely the  conservation laws of mass.

\section*{Acknowledgements}
The research  was supported by Russian Science Foundation Grant No 18-11-00238
`Hydrodynamics-type equations: symmetries, conservation laws, invariant difference schemes'.

\end{document}